\documentclass[preprint,12pt,authoryear,review]{elsarticle}

\usepackage{amssymb}
\usepackage{amsthm}
\usepackage{amsmath}
\usepackage{tabularx}
\usepackage{lineno}
\usepackage{booktabs}
\usepackage{textgreek}
\usepackage{float}

\usepackage{xcolor}

\usepackage{caption}
\usepackage{subcaption}

\begin{document}
\begin{frontmatter}

\title{Corneal deformation mapping and FE-based strain analysis via digital image correlation: biomechanical changes after CXL and laser refractive surgery}

\author[i3a]{Benedetta Fantaci}
\author[mat,inma]{Alejandro Frechilla}
\author[ubern]{Matteo Frigelli}
\author[ubern]{Philippe Büchler}
\author[ubern]{Sabine Kling}
\author[i3a,ciber]{Bego\~na Calvo}

\address[i3a]{I3A -- Instituto de Investigaci\'on en Ingenier\'ia de Arag\'on, Universidad de Zaragoza, Spain}
\address[mat]{Departamento de Ciencia y Tecnología de Materiales y Fluidos, Universidad de Zaragoza, Spain}
\address[inma]{Instituto de Nanociencia y Materiales de Aragón, INMA, CSIC-Universidad de Zaragoza, Spain}
\address[ubern]{ARTORG Center for Biomedical Engineering Research, University of Bern, Bern, Switzerland}
\address[ciber]{Centro de Investigaci\'on Biom\'edica en Red en Bioingenier\'ia, Biomateriales y Nanomedicina (CIBER-BBN), Spain.}


\begin{abstract}
\sloppy
Accurate assessment of corneal mechanical properties is critical for understanding ocular biomechanics, predicting refractive surgery outcomes, and optimizing cross-linking (CXL) treatments. Conventional uniaxial tensile test is limited by non-physiological boundary conditions and simplified stress distributions. Inflation testing more closely reproduces the in vivo stress state but has traditionally lacked full-field deformation mapping. In this work, we present an integrated experimental-computational protocol combining inflation testing of freshly enucleated porcine eyes with high-resolution three-dimensional digital image correlation (3D-DIC). Fifteen corneas were analyzed across three cohorts: (i) de-epithelialized controls, (ii) CXL-treated (standard Dresden protocol), and (iii) anterior stromal ablation via femtosecond laser. Samples were subjected to controlled intraocular pressure (IOP) elevations up to 40 mmHg. The 3D-DIC approach provided dense, pointwise displacement and strain maps across the anterior surface, successfully quantifying the localized stiffening effects of CXL and the increased compliance induced by stromal ablation. These full-field kinematic data were integrated into a membrane-theory finite element framework to resolve principal in-plane strains, that were used for subsequent inverse modeling to derive anisotropic hyperelastic parameters of porcine corneal tissue. Overall, the method establishes an end-to-end route from physiologic loading to full-field strain mapping and constitutive parameter identification, enabling quantitative evaluation of treatment-induced biomechanical changes in the cornea.
\end{abstract}


\begin{graphicalabstract}
\includegraphics[width=1\textwidth]{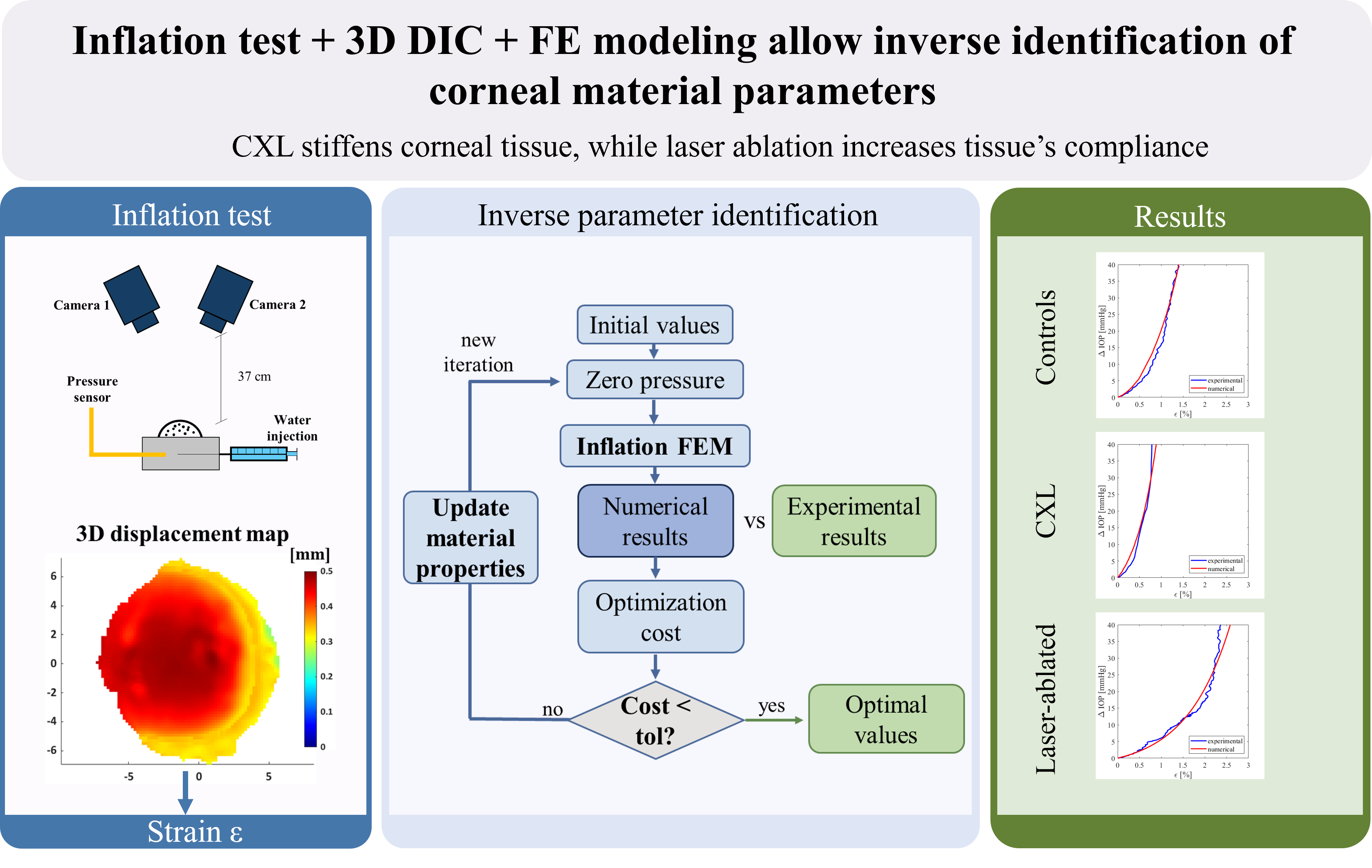}
\end{graphicalabstract}

\begin{highlights}
\item Inflation tests on porcine corneas were combined with 3D-DIC.
\item Three groups were analyzed: controls, CXL-treated, and laser-ablated corneas.
\item Full-field corneal surface strains were computed from 3D-DIC kinematics.
\item Inverse optimization identified GOH material parameters for each group.
\item CXL stiffened the tissue, whereas laser ablation increased compliance.
\end{highlights}

\end{frontmatter}

\section{Introduction}
The cornea plays a crucial role in maintaining the structural integrity and optical performance of the eye. The corneal stroma provides a transparent yet mechanically resilient structure that maintains corneal curvature and resists the loads generated by intraocular pressure (IOP), the fluid pressure exerted by the ocular humors. Due to its high transparency and the refractive-index contrast between the air and the cornea, it contributes nearly two thirds of the total refractive power of the eye. The cornea has a highly ordered stromal microstructure, with collagen lamellae arranged along preferential orientations \citep{Meek2015}. The mechanical response of the cornea is markedly nonlinear, as variations in fibrillar architecture, local hydration, and crosslinking confer anisotropic, strain-dependent stiffness.

Because corneal biomechanics affect visual function, characterizing corneal mechanical behavior is important for both clinical decision-making and computational modeling. Quantifying corneal material properties improves prediction of tissue responses to interventions and supports optimization of established procedures, such as refining CXL protocols and laser refractive surgery planning.
Direct in vivo identification of corneal material parameters remains challenging. Although non-contact deformation measurements (e.g., air-puff tonometry), OCT-based elastography, and Brillouin microscopy can probe corneal biomechanics in vivo, they typically yield metrics that are not straightforward to translate into constitutive parameters and may reflect the coupled response of the whole eye, (i.e., the combined effects of ocular geometry, IOP, and tissue material properties), making it difficult to isolate the intrinsic contribution of corneal tissue. Consequently, patient-specific constitutive parameter estimation is most often pursued using ex vivo mechanical tests (uniaxial, biaxial, or inflation) combined with inverse finite element (FE) analysis \citep{elsheikh2007,elsheikh2008, whitford2016,Ariza2017,nambiar2022, Frigelli2025}.
Although uniaxial and biaxial tests have historically been used to characterize corneal tissue, they may not reproduce the physiological three-dimensional loading imposed by IOP and can be affected by artifacts of specimen preparation (e.g. fiber severing) \citep{boyce2008,boschetti2012,wang2021review}. 

Unlike uniaxial and biaxial tests, inflation testing preserves corneal geometry and reproduces a more physiological loading state, enabling assessment of the pressure–deformation response. Nevertheless, many inflation studies rely on apex-only measurements or do not provide full-field displacement/strain mapping \citep{2Elsheikh2007,Wilson2020CornealBiomechanics}. To overcome these limitations, recent advances in optical metrology have employed three-dimensional digital image correlation (DIC), a non-contact technique that enables the measurement of full-field surface displacements and strains. DIC provides high-resolution displacement fields from stereo image pairs and has been successfully deployed in biomedical contexts \citep{wu2016}.

In ophthalmology, DIC has been used to quantify deformation in ocular tissues such as the sclera and cornea during inflation and quasi-static tests, enabling the extraction of robust mechanical biomarkers while mitigating rigid-body artifacts \citep{nguyen2010}. More recently, OCT-based digital volume correlation (DVC) has enabled depth-resolved full-field strain mapping during corneal inflation, revealing localized principal strains co-localized with stromal striae and predominantly oriented perpendicular to the striae \citep{Wu2024}. Beyond biological tissues, DIC has also been applied to characterize the mechanical behavior of medical devices, such as intraocular lenses \citep{cabeza2023}.
In inflation tests coupled with DIC, a random speckle pattern is applied to the corneal surface and recorded during pressurization using one or more synchronized cameras. The recorded images are then correlated through pattern-matching algorithms to reconstruct the three-dimensional deformation field with high spatial resolution. 
This approach allows for accurate quantification of displacement over the corneal dome, enabling computation of surface strain distributions and inverse identification of mechanical parameters when combined with FE analyses \citep{zhang2018,qiao2021}. Interpreting these full-field kinematic measurements in terms of intrinsic tissue behavior still requires computational models that explicitly account for corneal geometry, boundary conditions, and IOP-driven loading. In particular, FE models calibrated against inflation–DIC/DVC data can reconstruct the underlying stress state, support inverse estimation of constitutive parameters, and quantify regional mechanical changes induced by interventions (e.g., CXL or stromal ablation) under both physiological and altered conditions.

The present study introduces an integrated experimental–computational framework that combines ex vivo inflation testing of porcine corneas with full-field deformation mapping and inverse FE analysis to identify corneal material properties. The protocol is applied consistently across three clinically relevant conditions: de-epithelialized untreated controls,  CXL-treated corneas (standard Dresden protocol), and corneas subjected to femtosecond-laser anterior stromal ablation. High-fidelity pressure–deformation data are acquired under controlled inflation conditions, and an FE model replicating the experimental boundary conditions and loading is developed and validated against the measurements. The full-field kinematics are then used in an inverse analysis to estimate the anisotropic hyperelastic parameters of the Gasser–Ogden–Holzapfel constitutive model \citep{Gasser2006} that best reproduce the measured responses in each group. This unified identification pipeline enables quantitative comparison of treatment-induced changes in corneal mechanics and provides an analysis-ready basis for predictive simulations of corneal deformation under physiological and altered conditions.

\section{Materials and Methods}

\subsection{Sample preparation and grouping}

Fifteen freshly enucleated porcine eyes were obtained from a local slaughterhouse within two hours post-mortem and stored in sterile saline solution at 4\,$^\circ$C until use. All specimens were inspected under a surgical microscope to ensure the absence of epithelial damage, stromal opacities, or other visible defects. The experiments were conducted within 4 hours from enucleation to preserve the native biomechanical properties of the tissue.

The eyes were randomly assigned to three experimental groups (n=5 per group): (1) the control group, consisting of de-epithelialized, untreated corneas; (2) the CXL group, subjected to the standard Dresden corneal CXL protocol; and (3) the laser group, in which the anterior 350\,$\mu$m of the stroma were removed using a femtosecond ultraviolet (UV) laser system. 
All eyes were mounted in a custom 3D-printed ocular holder to maintain consistent alignment and curvature during treatment and subsequent mechanical testing (Figure \ref{inflation set up}).

\subsection{Corneal cross-linking (CXL) treatment}

The samples belonging to the CXL group underwent corneal cross-linking following the original \textit{Dresden Protocol} established by  \citep{seiler2003,iseli2008}. After epithelial debridement of the central 8\,mm zone, a riboflavin solution (0.1\% riboflavin-5-phosphate in 20\% dextran T-500) was applied to the corneal surface every five minutes for 30 minutes to ensure stromal saturation. Ultraviolet-A (UVA) irradiation was then delivered at a wavelength of 370\,nm, an irradiance of 3\,mW/cm$^2$, and a working distance of 1\,cm for 30 minutes, yielding a total energy dose of 5.4\,J/cm$^2$. During exposure, riboflavin drops were reapplied at five-minute intervals to maintain consistent hydration and photosensitizer concentration.

This protocol, which remains the clinical gold standard for corneal cross-linking, induces the formation of covalent bonds between collagen fibrils, increasing stromal stiffness and resistance to enzymatic degradation. Following treatment, the eyes were rinsed with balanced saline solution and maintained in a humidified environment until testing.

\subsection{Laser ablation treatment}
Partial stromal ablation was performed using a femtosecond laser system (Carbide CB3-40W+CBM03-2H-3H, Light Conversion, Lithuania) operating at a wavelength of 343\,nm (third harmonic, UV region) for the laser group. The beam exhibited an elliptical Gaussian intensity profile with $1/e^{2}$ intensity diameters (defined at the radial location where the intensity decreases to $I_0/e^{2}$, with $I_0$ denoting the peak intensity and $e$, Euler's number) of approximately 64\,$\mu$m along the major axis and $0.89\times 64 \approx 57$\,$\mu$m along the minor axis. Here, the aspect ratio is defined as $D_{\mathrm{minor}}/D_{\mathrm{major}}=0.89$, where $D_{\mathrm{major}}$ and $D_{\mathrm{minor}}$ denote the corresponding $1/e^{2}$ diameters along the major and minor axes, respectively. The pulse duration was 238\,fs, and the laser was linearly polarized. 

\begin{figure}[h]
\centering
\includegraphics[width=0.9\textwidth]{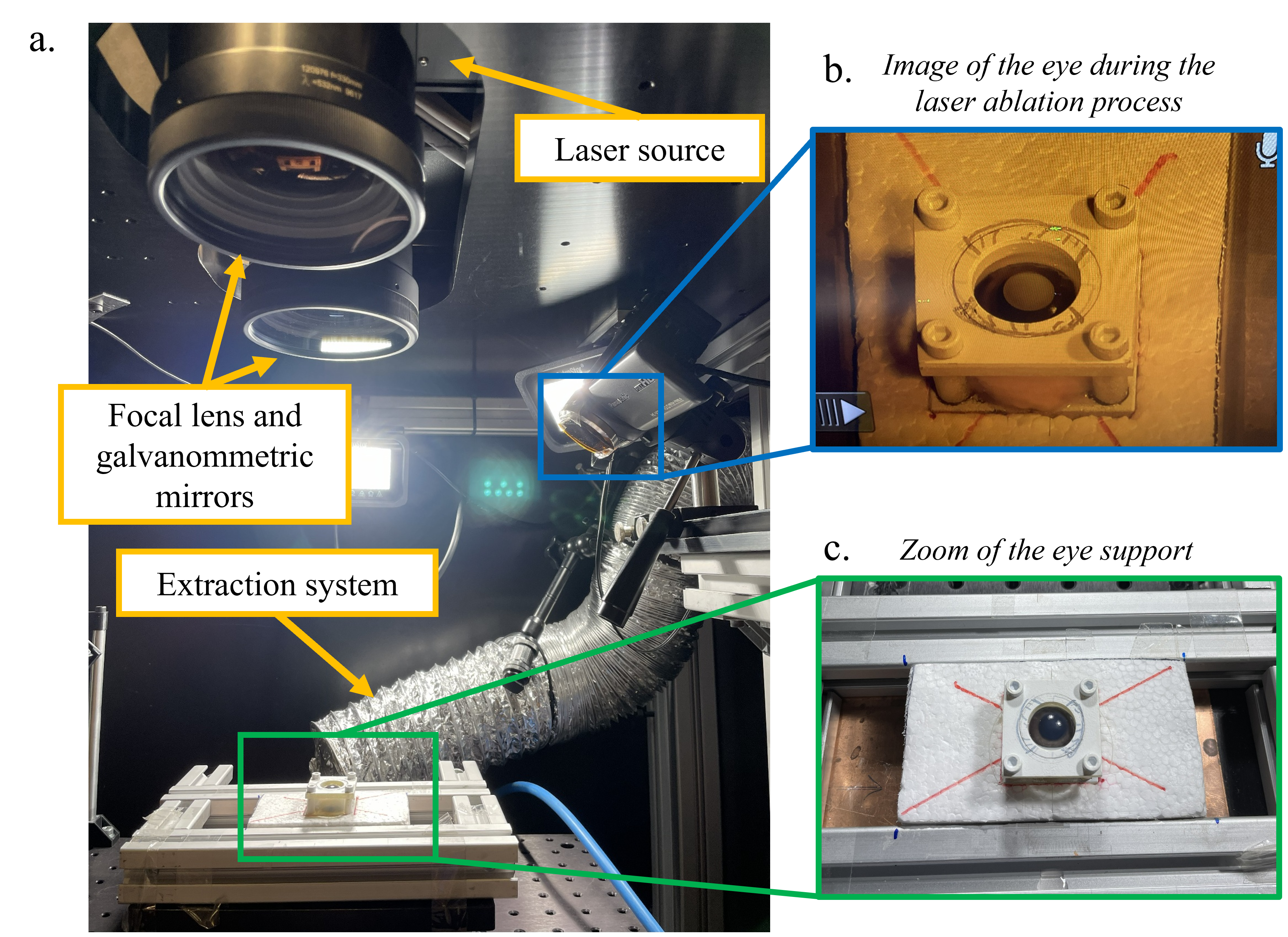}
\caption{\textbf{Experimental setup for the laser ablation process.} a. view of the laser system and the experimental set up prepared for the laser experiments; b. close-up view of the eye during the laser processing; c. close-up view of the eye inside the eye holder.}
\label{laser set up}
\end{figure}

Ablation was performed in ambient air using a custom-designed, 3D-printed eye holder that ensured stable fixation and reproducible positioning of each specimen (Figure~\ref{laser set up}). Laser parameters were selected to achieve controlled stromal removal while minimizing thermal accumulation (average power 3.3 W, pulse repetition rate 20 kHz, scanning speed 100 mm/s, and line spacing 10 $\mu$m). No corneal applanation (flattening) was applied. Importantly, the femtosecond UV laser system was employed as an experimental ablation platform and does not replicate clinical excimer-laser delivery strategies (e.g., wavefront-optimized or topography-guided ablation profiles). 

A uniform ablation pattern was delivered over a circular optical zone of 6.5 mm in diameter, centered on the corneal apex, with the intended removal depth prescribed in the laser control software. The corneal surface was irradiated by translating the laser beam to generate parallel ablation lines across the transverse plane, with adjacent lines separated according to the selected line spacing, thereby ensuring homogeneous coverage of the optical zone. The same scanning sequence was repeated three times to obtain a cumulative stromal removal of approximately 270 $\mu$m, as quantified by confocal microscopy (2300 $Pl\mu$ Sensofar, Terrassa, Spain). 
Under these conditions, ablation proceeded efficiently, and no evidence of collateral thermal damage was observed. The resulting surface morphology was qualitatively comparable to that reported in corneal photoablation procedures.

\subsection{Eye preparation and inflation testing}
\begin{figure}[h]
\centering
\includegraphics[width=0.9\textwidth]{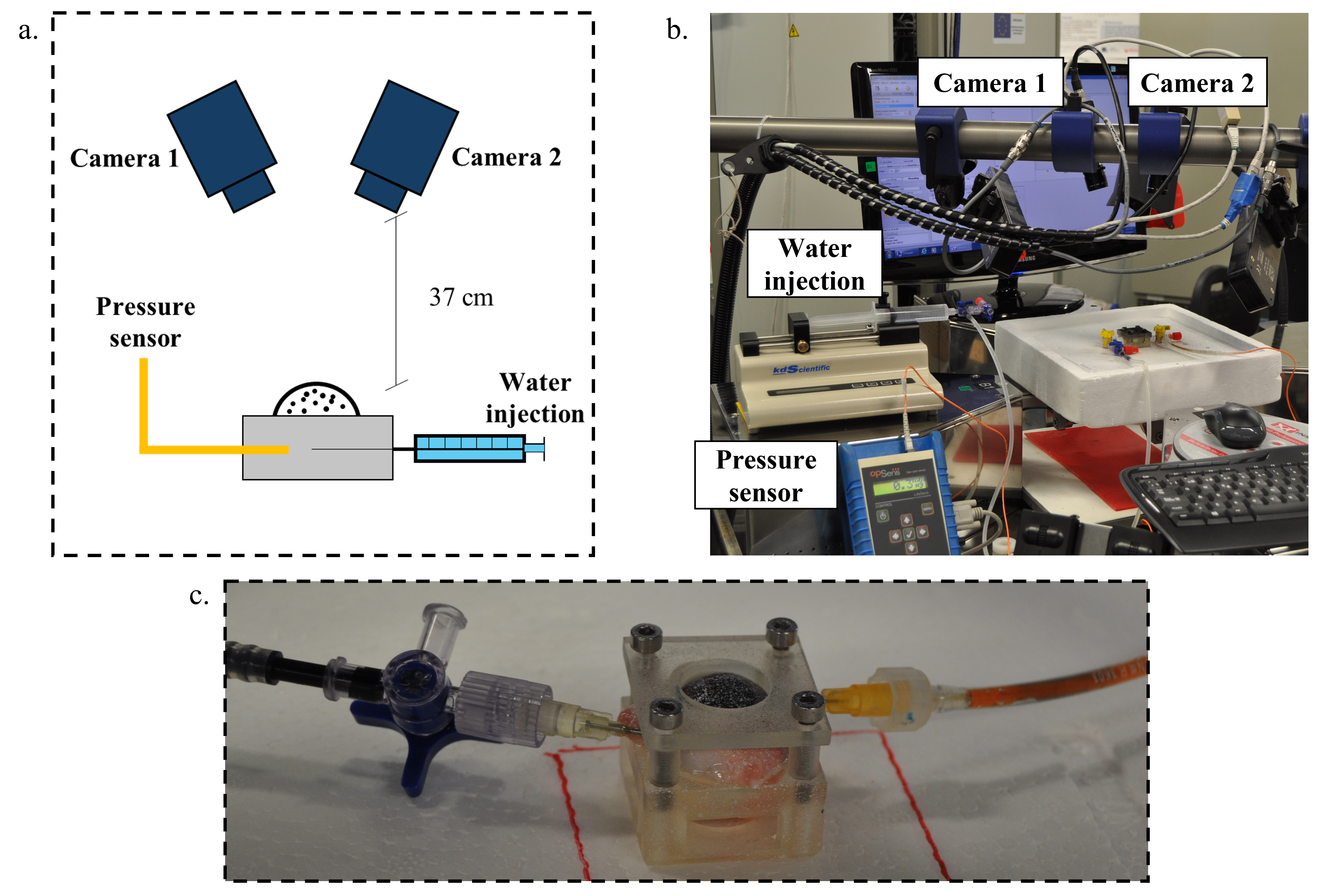}
\caption[Experimental setup for corneal inflation testing]{\textbf{Experimental setup for corneal inflation testing.} a. Schematic of the inflation chamber and optical arrangement; b. experimental setup; c. detail of the pressurization system with pressure and saline inlets; d. close-up view of the cornea during inflation.}
\label{inflation set up}
\end{figure}

All treatments were performed under standardized environmental conditions (22–24\,$^\circ$C, relative humidity 45–55\%) to minimize variability in corneal hydration. The control group underwent epithelial removal only, ensuring comparable surface exposure to hydration and mechanical testing conditions. Following treatment (i.e., CXL or laser ablation), all samples were immediately transferred to the inflation testing setup for biomechanical characterization.
The experimental setup  (Figure \ref{inflation set up}) consisted of:  
i) a pressure sensor inserted into the anterior chamber;  
ii) a precision syringe pump;  
iii) two cameras; 
iv) a custom-designed 3D-printed chamber, where the eyes were placed to preserve natural geometry and alignment along the optical axis relative to the imaging system.

Saline solution was infused into the posterior chamber of the eye through a 25G cannula connected to a precision syringe pump (Harvard Apparatus, USA) at a constant flow rate of 0.0833 ml/min. Intraocular pressure was continuously monitored via a calibrated pressure transducer (Honeywell 26PC series) inserted into the anterior chamber and connected to a digital acquisition system. 

The initial IOP measured immediately after insertion of the pressure sensor was recorded, yielding a mean value of 12 mmHg. The sensor was then reset to 0 mmHg, and the IOP was progressively increased from baseline (0\,mmHg) to 40\,mmHg.

The inflation process was recorded using a stereoscopic DIC system (Imager E-lite 2M, LaVision, Germany) comprising two cameras with a spatial resolution of 1280\,$\times$\,1024\,pixels operating at 3\,fps and a desktop computer with a Quad-core processor. The cameras were positioned at approximately 37\,cm from the corneal apex with an inter-camera separation of 13\,cm and fitted with identical 200\,mm f/4 Nikon lenses, providing a 30\textdegree observation angle. The two cameras were internally synchronized with LaVision software and high-power light-emitting diodes (LEDs) fed with distensibility coefficient (DC) to avoid flickering illuminated the sample. Uniform LED illumination was employed to prevent reflections and ensure consistent speckle visibility. Pressure and image acquisition were synchronized throughout all tests.

DIC requires the presence of a random pattern on the specimen to be able to record corneal displacement during the test. Prior to testing, the anterior corneal surface was coated with a random black speckle pattern using spray paint. Calibration of the stereo system was performed with a grid target following the DaVis (LaVision) protocol, before conducting the inflation tests. 

Images were acquired at 3 frames per second (fps). Image sequences were stored and processed to extract full-field displacements of the correlated speckle pattern. The data contained the reference coordinates (\textit{x} and \textit{y}, but not \textit{z}) and the 3D displacement (\textit{u\textsubscript{x}}, \textit{u\textsubscript{y}} and \textit{u\textsubscript{z}}) of the correlated speckle pattern. They were subsequently exported to MATLAB (R2024a, MathWorks, USA) and analyzed using the PIVMat 4.20 toolbox to reconstruct the three-dimensional displacement fields and compute surface deformations across the central corneal region. The resulting point cloud was meshed and the principal in-plane strains were calculated using an in-house Matlab code, which exploits the membrane theory of FEM ~\citep{rama2016} (see Subsection \ref{sec:strains}). 

As the pressure sensor was not synchronized with the cameras, the onset of the experiment was defined as the instant at which the pressure began to increase ($\Delta IOP > 1 $ mmHg).

\subsection{Strain calculation}
\label{sec:strains}
The displacement field obtained from DIC was post-processed to derive surface strains on the anterior cornea, in order to remove rigid-body translations from scleral motion and account for the zero initial positions assigned by DIC.

\begin{figure}[h]
\centering
\includegraphics[width=0.9\textwidth]{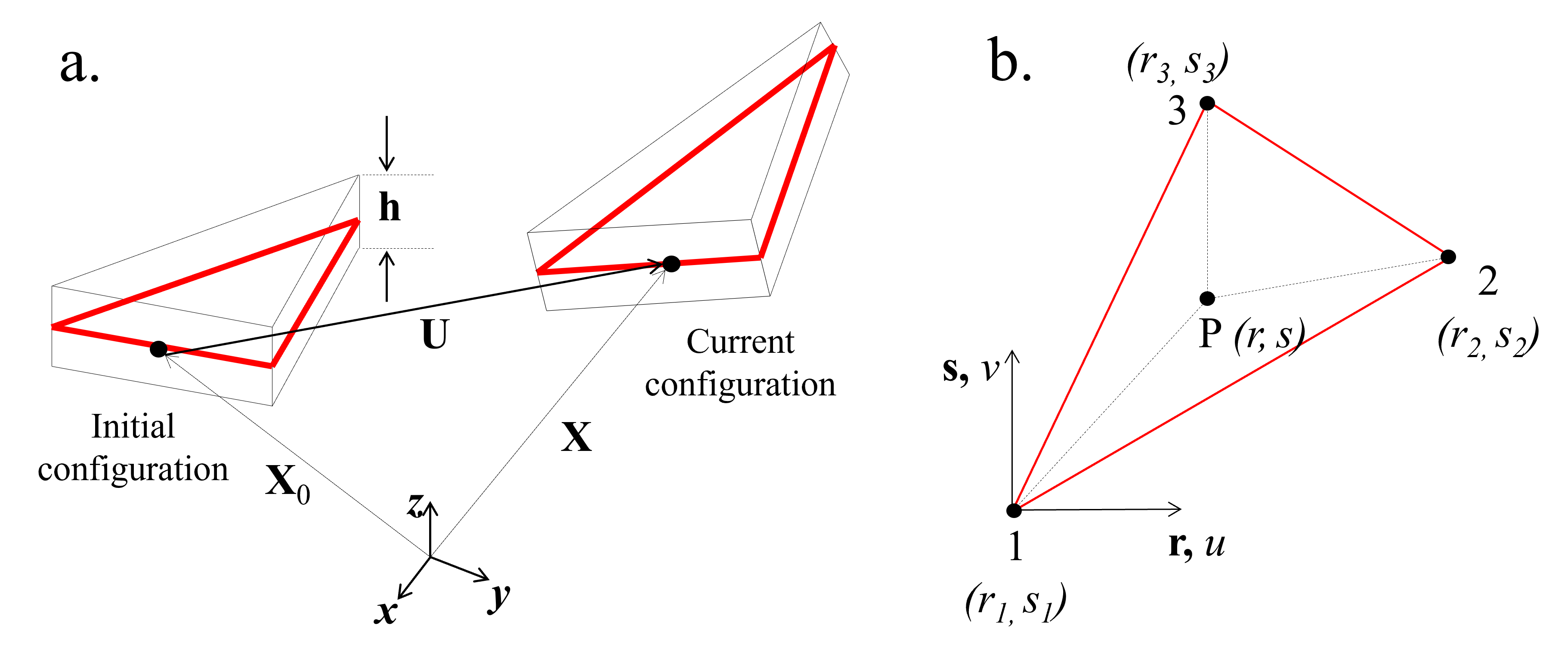}
\caption[Finite element strain computation]{\textbf{Finite element strain computation.} Workflow for deriving corneal surface strains from DIC-derived displacement fields using a co-rotational FE formulation. a. shell kinematics; b. Constant Strain
Triangular (CST) element \citep{rama2016}}
\label{Model Equations}
\end{figure}

The extracted 3D point cloud, typically comprising $\sim$1000 correlated surface nodes, was meshed using Delaunay triangulation to generate discrete triangular elements describing the surface geometry (Figure \ref{Model Equations}.a).

A local coordinate system ($r$, $s$, $t$) was defined for each element, with $r$ and $s$ lying on the element plane and $t$ being the normal to the element surface. As only the displacement of the corneal anterior surface was tracked by DIC, the thickness component through ($h$) was neglected in the following calculations, resulting in a planar strain field. Nodal displacements ($u$, $v$, $w$) were expressed in the local reference frame and used to compute in-plane strains via a co-rotational finite element approach (Figure \ref{Model Equations}b).

\sloppy
The method follows the formulation described by ~\citep{rama2016}, in which rigid-body rotations are decoupled from local deformations by assigning each triangular element its own co-rotational reference frame. Linearized strain–displacement relationships are then evaluated within this framework, assuming infinitesimal strains but finite rotations. The local in-plane strain field $\boldsymbol{\varepsilon} = \varepsilon_m + z \kappa_b$, containing contributions from the membrane strain ($\varepsilon_m$) and the flexural or bending strain ($z\kappa_b$), is given by:

\begin{equation}
\boldsymbol{\varepsilon}_m = \mathbf{B}_m \, \mathbf{d}_e, \quad
z \boldsymbol{\kappa}_b = z \, \mathbf{B}_b \, \mathbf{d}_e
\end{equation}

where $\mathbf{B}_m$ and $\mathbf{B}_b$ are the strain–displacement matrices, $\mathbf{d}_e$ the nodal displacement vector, and $z$ the distance from the mid-surface, which was neglected in this study as no information from the corneal thickness could be recorded. Consequently, the local strain field was approximated as the membrane strain:

\begin{equation}
\boldsymbol{\varepsilon} \approx \boldsymbol{\varepsilon}_m = \mathbf{B}_m \mathbf{d}_e \
\end{equation}

Once the local strain tensor was computed at each element, the corresponding maximum principal strain (MPS) was extracted to obtain a scalar strain metric independent of the local reference frame. The resulting MPS maps provided a spatially resolved quantification of corneal deformation under increasing IOP, enabling comparisons among CXL-treated, laser-ablated, and untreated control corneas.

The \textDelta IOP–strain curve for each experiment was obtained by computing the MPS over a circular region with a diameter of 5 mm. 

The same algorithm was used to compute surface MPS in the porcine FE model, thereby ensuring comparability for the inverse-analysis optimization.

\subsection{Cornea finite element model}

\begin{figure}[h]
\centering
\includegraphics[width=0.8\linewidth]{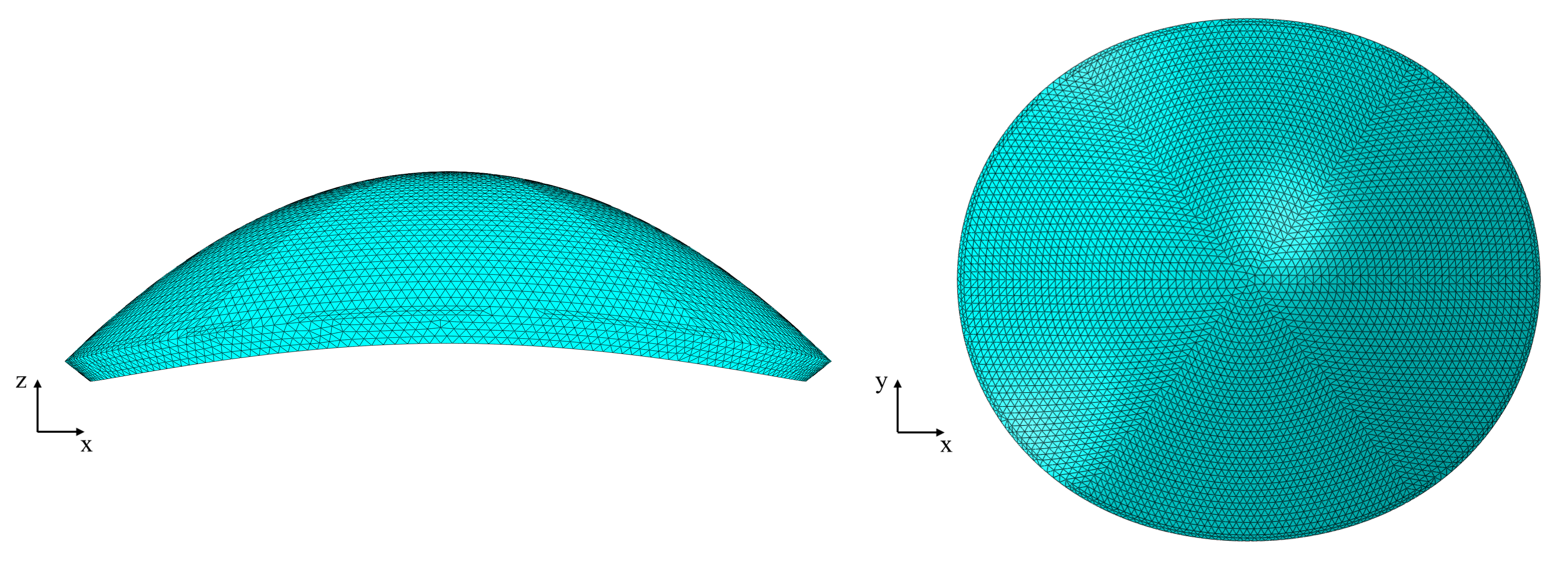}
\caption{\textbf{FE model of the porcine cornea.}}
\label{fig:fe_model}
\end{figure}

A FE model of the porcine cornea was developed to simulate the inflation response under controlled IOP increments.
Unlike the human cornea, which is approximately circular in anterior view, the porcine cornea exhibits a marked elliptical geometry.
Accordingly, both the anterior and posterior corneal surfaces were described using a biconic formulation expressed in elliptical coordinates (Figure~\ref{fig:fe_model}).
The corneal elevation was defined as
\begin{equation}
z(\theta) =
z_0 -
\frac{r(\theta)^2\,A(\theta)}
{1 + \sqrt{\,1 - r(\theta)^2\,B(\theta)\,}},
\label{eq:z_surface}
\end{equation}
where $z_0$ denotes the apex position and $\theta$ is the angular coordinate in the elliptical coordinate system. In Eq.~\eqref{eq:z_surface}, $r(\theta)$ is the in-plane radial coordinate, while $A(\theta)$ and $B(\theta)$ incorporate directional curvature and asphericity, respectively, as detailed below.

The in-plane elliptical geometry was defined as
\begin{equation}
\begin{aligned}
x(\theta) &= a \cos\theta, \\
y(\theta) &= b \sin\theta,
\end{aligned}
\label{eq:ellipse}
\end{equation}
where $a$ and $b$ are the semi-axes of the ellipse. The corresponding radial coordinate was computed as
\begin{equation}
r(\theta) = \sqrt{x(\theta)^2 + y(\theta)^2}.
\label{eq:radius}
\end{equation}

An angular offset with respect to the steepest meridian was introduced as
\begin{equation}
\Delta\theta = \theta - \theta_s,
\label{eq:delta_theta}
\end{equation}
where $\theta_s$ denotes the orientation of the steepest corneal meridian.

Directional curvature and asphericity were incorporated through
\begin{equation}
A(\theta) =
\frac{\cos^2\!\Delta\theta}{R_s}
+
\frac{\sin^2\!\Delta\theta}{R_f},
\label{eq:A_theta}
\end{equation}
and
\begin{equation}
B(\theta) =
(Q_s + 1)\frac{\cos^2\!\Delta\theta}{R_s^{2}}
+
(Q_f + 1)\frac{\sin^2\!\Delta\theta}{R_f^{2}},
\label{eq:B_theta}
\end{equation}
where $R_s$ and $R_f$ are the radii of curvature along the steepest and flattest meridians, respectively, and $Q_s$ and $Q_f$ are the associated asphericity coefficients.

Since individual corneal dimensions could not be measured for each tested eye, average values reported by \citet{menduni2018} were used.

Corneal tissue is made up mainly of the stroma, which represents~90\% \ of its thickness. The stroma consists of collagen fibrils embedded in the extracellular matrix (ECM), which confer a nonlinear, hyperelastic, anisotropic behavior to the tissue. 
In the central porcine cornea, stromal collagen shows weak in-plane preferential alignment (high dispersion) with a lamellar “crisscross” arrangement rather than a consistent global meridional axis, whereas toward the periphery/limbus collagen becomes increasingly tangential (circumferential) \citep{Hayes2007,elsheikh2008,Nguyen2014,Eltony2022,HatamiMarbini2025}.
To model the behavior of corneal tissue, the following strain energy density function was selected \citep{Gasser2006}:
\begin{equation}
U = C_{10}(\overline{I}_1 - 3)
+ \frac{1}{D} \left( \frac{(J^{el})^2 - 1}{2} - \ln J^{el} \right)
+ \frac{k_1}{2k_2} \sum_{\alpha=1}^{2} \left\{ \exp \left[ k_2 (\overline{E}_\alpha)^2 \right] - 1 \right\}.
\end{equation}
with
\begin{equation}
\overline{E}_{\alpha} 
= \kappa (\overline{I}_1 - 3) 
+ (1 - 3\kappa)(\overline{I}_{4(\alpha\alpha)} - 1).
\end{equation}
where $C_{10}$ is the material constant that controls the stiffness of the extracellular matrix, $k_{1}$ and $k_{2}$ are constants related to the stiffness of the fibers and the non-linear behavior of the fibers, respectively; $k$ represent the fiber dispersion; $\alpha$ is the number of families of fibers; $\overline{I}_1$ is the first invariant of $\mathbf{\overline{C}}$; $J_{el}$ is the jacobian and $\overline{I}_{4(\alpha\alpha)}$ are the pseudo-invariants of $\mathbf{\overline{C}}$ and $\mathbf{A}_\alpha$.

To reproduce experimental observations, two orthogonal fiber families were defined in the central stroma and modeled assigning intermediate dispersion $k=0.17$ to represent a weakly anisotropic in-plane response; toward the limbus, fibers were assumed to progressively align circumferentially.

The model was meshed with quadratic tetrahedral elements and boundary conditions (BCs) were applied at the limbus, restricting the nodes to radial displacements only within a spherical coordinate system centered at the origin of the best-fit sphere to the corneal geometry. The application of proper BCs allowed us to exclude the sclera from the FE model, avoiding additional uncertainty associated with its poorly characterized mechanical properties and reducing computational cost.

For loading conditions, the IOP was modeled as a uniform pressure load applied to the posterior corneal surface. Since the eye is under physiological IOP during topographic acquisition, the measured geometry corresponds to a deformed configuration, with the actual undeformed shape unknown \citep{Elsheikh2013}. We therefore employed an iterative algorithm, following \citep{ ariza2018}, to recover the unknown stress-free reference geometry prior to simulating the inflation tests.

A mesh sensitivity analysis was conducted to balance the model accuracy and computational cost. Three mesh sizes (0.15, 0.20, and 0.25 mm) were tested by simulating an inflation up to 40 mmHg; the apical displacement was extracted for each mesh and compared with that of the finest mesh. After selecting an element size of 0.20 mm, the inflation test was simulated within an optimization framework (see Section~\ref{sec:optimization}).

The simulation consisted of two steps: in the first step, the eye was pressurized with a physiological IOP of 12 mmHg, as detected during the experiments; in the second step, the inflation test was simulated, by incrementally applying an additional pressure up to 40 mmHg.

The effect of CXL was modeled via a stiffening factor $K_{\mathrm{CXL}}$ applied to constants $k_1$ and $k_2$. Depth dependence was incorporated by specifying a linear reduction of $K_{\mathrm{CXL}}$ with stromal depth, such that $K_{\mathrm{CXL}}=1$ at $\mathrm{depth}_{\mathrm{CXL}} = 300\, \mu m$. This depth was selected based on the mean demarcation line depth observed in keratoconus following the Dresden protocol \citep{Frigelli2025}.

$K_{CXL}$ was defined according to the following equation: 
\begin{equation}
\resizebox{\textwidth}{!}{$
F_{\mathrm{CXL}} =
\Bigl[
  (K_{\mathrm{CXL}}-1)\, \mathrm{s}
  - (K_{\mathrm{CXL}}-1)\!\left(\frac{\mathrm{CCT}-\mathrm{depth}_{\mathrm{CXL}}}{\mathrm{CCT}}\right)
  + \frac{\mathrm{depth}_{\mathrm{CXL}}}{\mathrm{CCT}}
\Bigr]
\left(\frac{\mathrm{CCT}}{\mathrm{depth}_{\mathrm{CXL}}}\right)
$}
\end{equation}
where $F_{\mathrm{CXL}}$ denotes the point-wise linear factor used to scale the material constant; a local through-thickness coordinate $s$ was defined, such that $s=0$ at the anterior surface and $s=1$ at the posterior surface.

To simulate inflation in laser-ablated eyes, we reduced the corneal thickness by 270 \textmu m, assuming that the ablation does not alter stromal material properties. Consequently, the change in mechanical behavior is attributed solely to the decreased thickness, which plays a key role in corneal mechanical stability.

\subsection {Optimization Algorithm for Material Identification}
\label{sec:optimization}
\begin{figure}[h]
\centering
\includegraphics[width=0.8\linewidth]{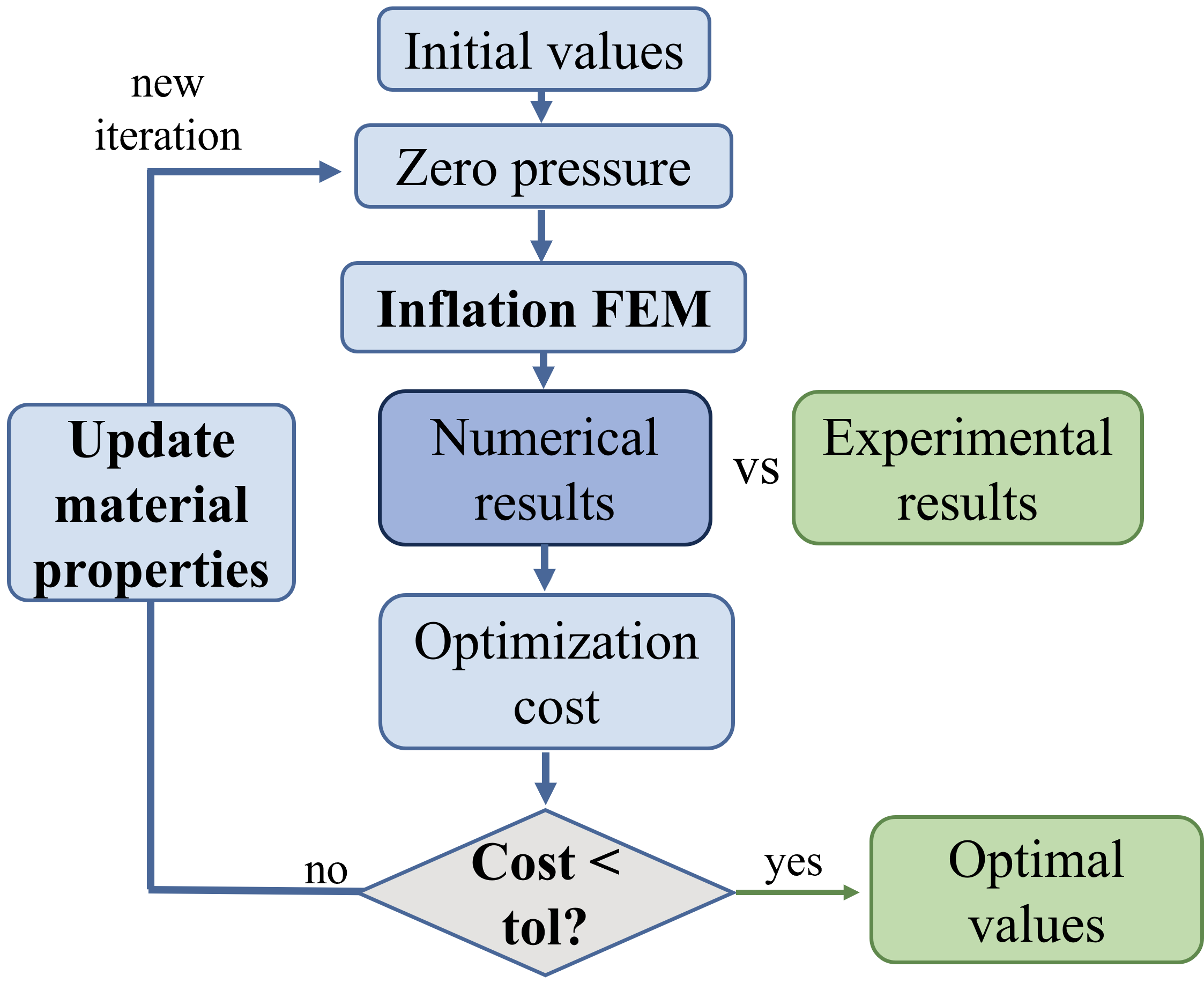}
\caption{\textbf{Flowchart of the optimization framework used to identify corneal material properties from inflation tests.}}
\label{fig:optimization_flow}
\end{figure}

An iterative optimization process was followed to compute the corneal material properties for the control and cross-linked corneas (Figure~\ref{fig:optimization_flow}).

The procedure begins by assigning a first set of material parameters to simulate the inflation test. The initial material properties were taken from the fitting performed by ~\citet{Pandolfi2008} on the experimental inflation tests of porcine \citep {Anderson2004}. The retrieved values were $C_{10} = 0.015\,\text{MPa}$, $k_{1} = 0.02\,\text{MPa}$ and $k_{2} = 400$. For the dispersion parameter $k$, we selected the intermediate value 0.17 between 0 (completely anisotropic) and 0.33 (completely isotropic), and it was kept fixed throughout the optimization procedure.

Before running the inflation simulation, the zero-pressure configuration of the cornea is computed. The inflation simulation is then performed, producing numerical displacement fields that are used to calculate surface deformations, as described in Section \ref{sec:strains}. For both experiments and simulations, the strains are computed on a circular area with a radius of 5 mm, and their average value is obtained.

The discrepancy between experimental and numerical mean strains was quantified through an $L_2$-based cost function.
Let $\bar{\varepsilon}^{\,\mathrm{exp}}_i$ and $\bar{\varepsilon}^{\,\mathrm{sim}}_i(\boldsymbol{\theta})$ denote the experimental and simulated mean strains at the $i$-th loading level, respectively, with $\boldsymbol{\theta}$ the vector of material parameters. The residual vector was defined as
\begin{equation}
\mathbf{r}(\boldsymbol{\theta}) =
\bigl[\bar{\varepsilon}^{\,\mathrm{exp}}_1-\bar{\varepsilon}^{\,\mathrm{sim}}_1(\boldsymbol{\theta}),\;
\dots,\;
\bar{\varepsilon}^{\,\mathrm{exp}}_N-\bar{\varepsilon}^{\,\mathrm{sim}}_N(\boldsymbol{\theta})\bigr]^{\mathsf{T}},
\end{equation}
where $N$ is the number of loading levels considered (after interpolating the simulated response onto the experimental sampling points when needed). The cost function was computed exactly as
\begin{equation}
J(\boldsymbol{\theta}) = 100\,\frac{\left\|\mathbf{r}(\boldsymbol{\theta})\right\|_2}{N},
\label{eq:cost_function}
\end{equation}
with $\|\cdot\|_2$ denoting the Euclidean norm. The factor $100$ scales the mismatch to percentage units, while division by $N$ normalizes the metric by the number of points.
The optimization was terminated when $J(\boldsymbol{\theta})<1$, corresponding to a mismatch below $1\%$ according to the definition in Eq.~\eqref{eq:cost_function}.

For the controls, the constants $C_{10}$, $k_{1}$, and $k_{2}$ were optimized. For CXL corneas, only the parameter $K_{\mathrm{CXL}}$ was adjusted using the material properties obtained from control optimization. Finally, for laser-ablated corneas, no optimization was performed: the thickness of the porcine model was reduced, as described above, and a simulation using the same material properties as the controls was run.

All simulations were performed in Abaqus/Standard 2022 (Dassault Systèmes, France).

\subsection{Statistical analysis}

For each group (controls, CXL, laser), individual strain--pressure curves were interpolated onto a uniform pressure grid (0--40~mmHg, step 0.1~mmHg). Group mean curves and standard deviations were computed across specimens to obtain representative inflation responses.

Global differences in inflation behavior were assessed using a linear mixed-effects (LME) model fitted with MPS (\%) as the dependent variable and $\Delta$IOP, treatment group, and their interaction ($\Delta$IOP$\times$group) as fixed effects. Specimen-specific random intercepts and random slopes with respect to $\Delta$IOP were included to account for repeated measurements along each curve. For better interpretability, the fitted strain--$\Delta$IOP slopes ($d\varepsilon/d(\Delta\mathrm{IOP})$) were reported as their inverse, $d(\Delta\mathrm{IOP})/d\varepsilon$ (mmHg per \% MPS {$\varepsilon$}).

Group differences at specific loading levels were evaluated using model-based pairwise contrasts (control--CXL, control--laser, CXL--laser) derived from the LME at selected $\Delta$IOP values (20 and 40~mmHg); within each $\Delta$IOP level, $p$-values were adjusted for the three comparisons using the Holm correction. Statistical significance was set at $p<0.05$.

\section{Results}

\subsection{Statistical analysis of inflation curves}

\begin{figure}[ht]
\centering
\includegraphics[width=\textwidth]{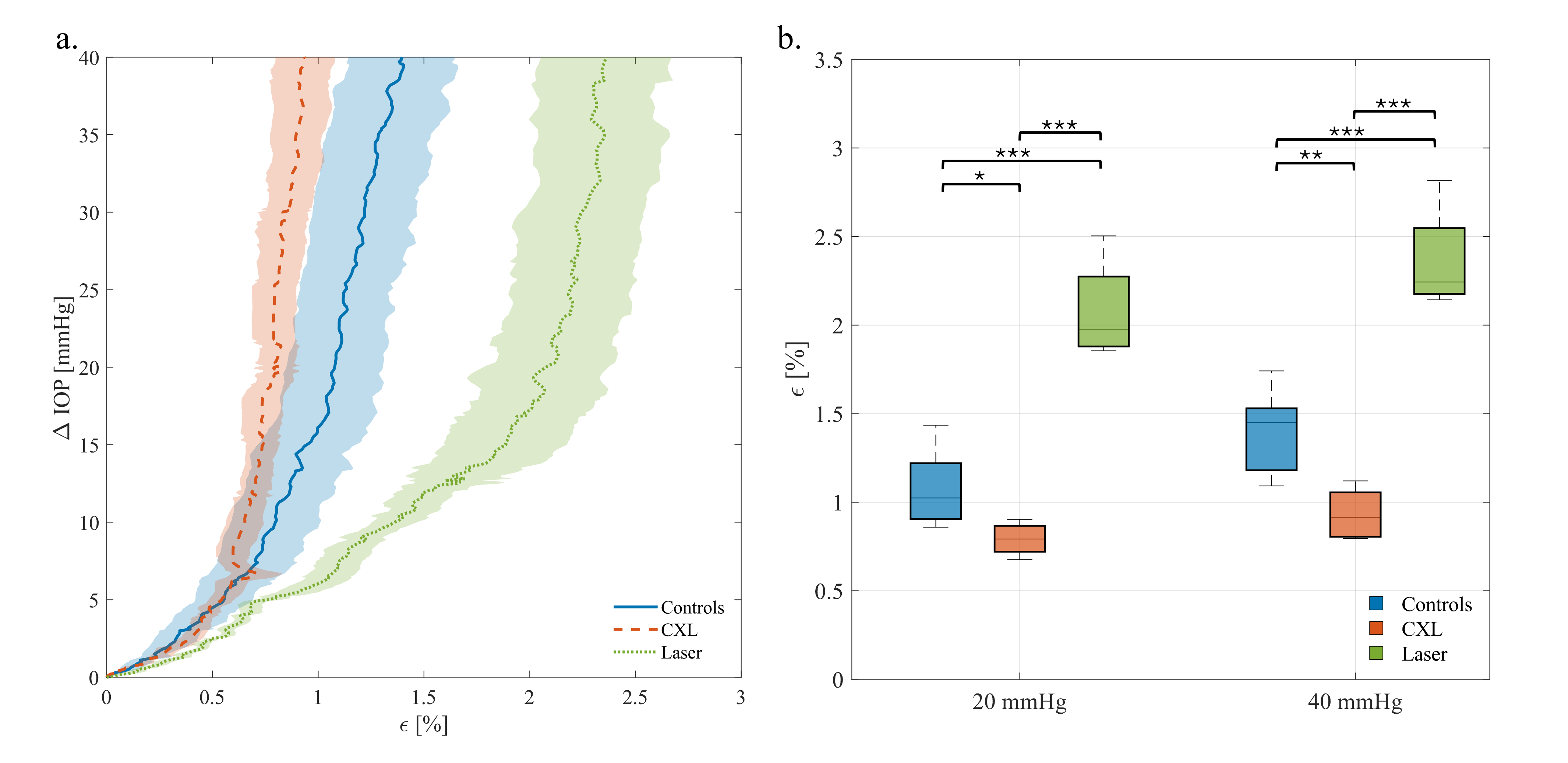}
\caption{\textbf{Experimental inflation response of porcine corneas.}
a. Mean inflation curves (mean $\pm$ SD) showing the relationship between MPS $\varepsilon$ and pressure increment $\Delta$IOP for control, CXL-treated, and laser-ablated corneas.
b. Boxplots of MPS $\varepsilon$ at $\Delta$IOP = 20 and 40~mmHg. Significance is indicated as * $p<0.05$, ** $p<0.01$, and *** $p<0.001$ (Holm-corrected).}
\label{fig:experimental_results}
\end{figure}

The inflation curves of control, CXL-treated, and laser-ablated porcine corneas are shown in Figure~\ref{fig:experimental_results}.a. 

Inflation responses differed across treatments (Figure~\ref{fig:experimental_results}.a-b). Strain increased with pressure ($p<0.001$), with significant effects of group ($p<0.001$) and pressure$\times$group interaction ($p<0.001$), indicating treatment-dependent strain--pressure relationships.

Using controls as reference, the inflation response was quantified in terms of the pressure--strain slope defined as $d(\Delta\mathrm{IOP})/d\varepsilon$ (mmHg per \% strain), i.e., the pressure increment required to produce 1\% strain (higher values indicate a stiffer response). For controls, the estimated slope was $38.7$~mmHg/\%{$\varepsilon$} (95\%~CI: $33.8$--$45.3$). The CXL group exhibited a higher pressure requirement per unit strain, with $d(\Delta\mathrm{IOP})/d\varepsilon = 73.8$~mmHg/\%{$\varepsilon$} (95\%~CI: $57.7$--$102.2$), consistent with a stiffer inflation behavior. In contrast, the laser group showed a lower pressure requirement per unit strain, with $d(\Delta\mathrm{IOP})/d\varepsilon = 21.1$~mmHg/\%{$\varepsilon$} (95\%~CI: $19.4$--$23.2$), consistent with a more compliant response. These trends are consistent with Figure ~\ref{fig:experimental_results}.a, where laser presents higher strains at a given $\Delta$IOP (right-shifted curve), whereas CXL yields lower strains as $\Delta$IOP increases.

Model-based pairwise contrasts (Holm-corrected within each pressure level) are reported in Table~\ref{tab:lme_contrasts} (see also Figure ~\ref{fig:experimental_results}.b). The laser-treated specimens exhibited higher MPS than both controls and CXL at all evaluated levels of $\Delta$IOP. CXL did not differ from controls at low loading (10~mmHg) but showed significantly lower strain from 20~mmHg onwards. The magnitude of between-group differences increased with $\Delta$IOP, consistent with the significant pressure$\times$group interaction.

\begin{table}[H]
\centering
\caption{Model-based pairwise contrasts in strain (\%) at selected loading levels. Positive differences indicate higher strain in the first group listed. Significance is indicated as * $p<0.05$, ** $p<0.01$, and *** $p<0.001$ (Holm-corrected within each $\Delta$IOP level).}
\label{tab:lme_contrasts}
\small
\setlength{\tabcolsep}{4pt}
\renewcommand{\arraystretch}{1.1}
\begin{tabular}{c l c l l}
\toprule
\shortstack{$\Delta$IOP\\(mmHg)} & Contrast & Diff (\%) & 95\% CI & $p_{\mathrm{Holm}}$ \\
\midrule
10 & CXL -- Control   & -0.121 & [$-0.270$, 0.028]  & $>0.05$ \\
10 & Laser -- Control &  0.570 & [0.412, 0.728]     & $<0.001^{***}$ \\
10 & Laser -- CXL     &  0.691 & [0.533, 0.849]     & $<0.001^{***}$ \\
\addlinespace
20 & CXL -- Control   & -0.244 & [$-0.438$, $-0.049$] & $<0.05^{*}$ \\
20 & Laser -- Control &  0.784 & [0.578, 0.991]     & $<0.001^{***}$ \\
20 & Laser -- CXL     &  1.028 & [0.822, 1.235]     & $<0.001^{***}$ \\
\addlinespace
30 & CXL -- Control   & -0.367 & [$-0.610$, $-0.124$] & $<0.01^{**}$ \\
30 & Laser -- Control &  0.999 & [0.741, 1.257]     & $<0.001^{***}$ \\
30 & Laser -- CXL     &  1.366 & [1.108, 1.623]     & $<0.001^{***}$ \\
\addlinespace
40 & CXL -- Control   & -0.490 & [$-0.783$, $-0.197$] & $<0.01^{**}$ \\
40 & Laser -- Control &  1.213 & [0.902, 1.524]     & $<0.001^{***}$ \\
40 & Laser -- CXL     &  1.703 & [1.392, 2.014]     & $<0.001^{***}$ \\
\bottomrule
\end{tabular}
\end{table}

Overall, the experimental data define three mechanically distinct responses that provide a robust basis for computational modeling and parameter identification.

\subsection{Experimental--numerical comparison and model performance}
\begin{figure}[ht]
\centering
\includegraphics[width=\textwidth]{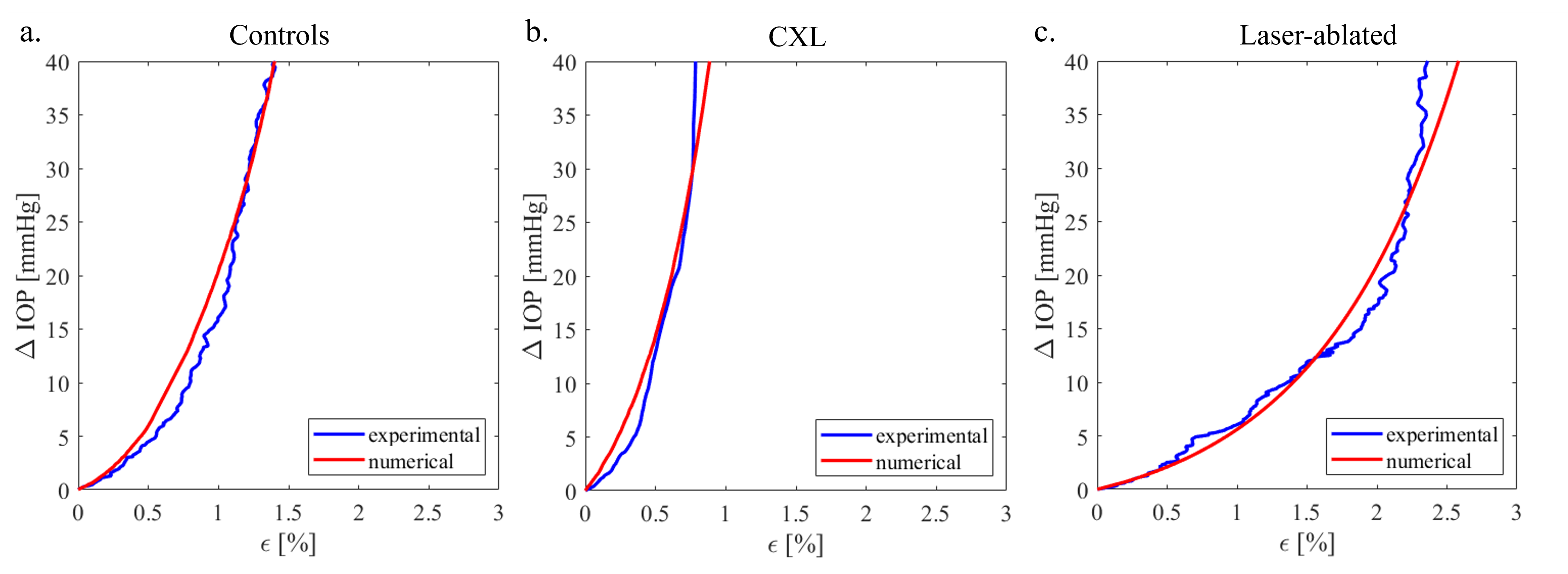}
\caption{\textbf{Comparison between experimental measurements and finite element simulations.}
Inflation curves showing pressure increment $\Delta$IOP as a function of  MPS $\varepsilon$ for: a. control corneas; 
b. CXL-treated corneas; 
c. laser-ablated corneas. \\
Experimental data are shown in blue, while FE simulations are shown in red.
The reported loss value quantifies the discrepancy between experimental and numerical responses used in the inverse optimization.}
\label{fig:exp_vs_sim}
\end{figure}

From the optimization process, the following material properties were obtained: for the control group, $C_{10}=0.002$ MPa, $k_1=0.0238$ MPa, and $k_2=583$ [-]; for the CXL-treated corneas, $K_{\mathrm{CXL}}=1.53$. For the laser-ablated corneas, the same material properties as those of the control group were used.
Figure~\ref{fig:exp_vs_sim} compares experimental inflation curves with FE simulations from each group. 
For control and CXL-treated corneas (Figure~\ref{fig:exp_vs_sim}.a-b), the FE model closely reproduced the experimental response across the full pressure range. 
This demonstrates that the adopted constitutive formulation and boundary conditions accurately capture the baseline biomechanics of untreated porcine corneas.

In laser-ablated corneas (Figure ~\ref{fig:exp_vs_sim}.c), the FE model captured the overall inflation behavior but showed some deviations in reproducing the tissue’s nonlinear response at higher strains ($\varepsilon > 1.5\%$). Nevertheless, the numerical prediction remained within the experimental standard deviation (see Figure ~\ref{fig:experimental_results}.a).

Overall, these results demonstrate that the adopted constitutive formulation and modeling approach accurately capture the biomechanics of porcine corneas under different conditions.

\section{Discussion}

This work presents an experimental--computational framework to quantify corneal biomechanics by combining whole-eye inflation testing with full-field 3D DIC and strain computation via a membrane-theory FE post-processing pipeline. By employing full-field kinematics rather than apex-only displacement, the approach enables robust characterization of inflation responses, while mitigating rigid-body motion artifacts, and provides critical constraints for inverse identification of constitutive parameters and biomechanical modeling.

Porcine corneas have been widely used as human analogs due to their comparable geometry and collagen organization, and because they are readily available whereas human corneas are difficult to obtain. Previous studies have characterized their short- and long-term behavior under inflation, creep, and stress-relaxation conditions, revealing strong nonlinear stiffening with increasing pressure \citep{elsheikh2007,elsheikh2008,boschetti2012,chang2020}.
Nevertheless, quantitative determination of porcine corneal parameters through systematic inverse analysis remains limited, and reported data exhibit substantial variability due to differences in experimental protocols and constitutive assumptions.

The inflation responses revealed three distinct mechanical responses. Corneas treated with CXL exhibited a reduced strain--pressure slope compared with controls, with group differences becoming more pronounced as pressure increased, consistent with a stiffening effect that manifests most clearly in the nonlinear deformation regime. In contrast, laser-ablated corneas displayed an overall upward shift in strain at comparable \(\Delta \mathrm{IOP}\), indicating a more compliant response and reduced load-bearing capacity under the same pressure increments. These observations were supported by the statistical analysis, including a significant pressure$\times$group interaction in the mixed-effects model and model-based contrasts at representative loading levels.

Consistent with prior porcine cornea data, the measured inflation behavior was distinctly nonlinear, with an early transition from a low-stiffness regime to a collagen-dominated regime. In our mixed-effects analysis, group separation was captured by the  slope \(d(\Delta \mathrm{IOP})/d\varepsilon\) (mmHg per \% MPS {$\varepsilon$}), which was 38.7~mmHg/\%{$\varepsilon$} (95\% CI: 33.8--45.3) for controls, increased to 73.8~mmHg/\%{$\varepsilon$} (95\% CI: 57.7--102.2) after CXL, and decreased to 21.1~mmHg/\%{$\varepsilon$} (95\% CI: 19.4--23.2) after laser ablation, with a significant pressure$\times$group interaction indicating that between-group differences grew with increasing load. Notably, the CXL effect was negligible at low loading (\(\Delta \mathrm{IOP}=10\)~mmHg) but became significant from 20~mmHg onward, whereas laser-treated corneas showed higher strains than both other groups across all evaluated \(\Delta \mathrm{IOP}\) levels. This pattern agrees with inflation studies reporting a marked stiffness change around \(\sim 10\)~mmHg and steeper behavior beyond that point: Boschetti et al.\ documented a clear bilinear pressure--apex displacement response with an evident slope change at \(\sim 10\)~mmHg \citep{boschetti2012}, and Elsheikh et al.\ similarly reported that the transition in porcine corneas is comparatively abrupt and occurs around \(\sim 10\)~mmHg \citep{elsheikh2008}.

A mechanistic interpretation is that CXL increases stromal resistance to deformation by introducing additional intermolecular cross-links within the collagen network, effectively shifting the cornea toward a stiffer response at higher strains. Conversely, stromal ablation decreases stiffness primarily through thickness reduction and removal of load-bearing anterior tissue, yielding larger strains across the tested pressure range.

From the inverse FE identification, the optimized control parameters (\(C_{10}=0.002\)~MPa, \(k_1=0.0238\)~MPa, \(k_2=583\)) and the CXL stiffening factor (\(K_{\mathrm{CXL}}=1.53\) applied to the fiber-related constants) enabled the model to closely reproduce the experimental inflation response for controls and CXL-treated corneas across the tested pressure range, suggesting that the adopted constitutive assumptions, boundary conditions, and strain post-processing pipeline are collectively adequate to capture baseline and stiffened responses under the present experimental configuration. Importantly, the magnitude of the inferred CXL stiffening is consistent with intact-globe inflation data: Chang et al.\ reported \(\sim 28\%\) lower apex displacement after Dresden CXL at 27.25~mmHg and showed modulus ratios remaining approximately 1.4--1.5 throughout inflation, supporting a moderate but clear stiffening under physiological-to-mildly elevated IOP \citep{chang2020}. 
For the laser group, the model captured the overall trend but showed larger deviations at higher strains, with predictions remaining within the experimental standard deviation. 
Comparisons with uniaxial-strip studies should be interpreted cautiously: Boschetti et al.\ reported uniaxial tangent slopes of \(E_1 \approx 3.19\)~MPa and \(E_2 \approx 41.81\)~MPa, while inflation-based shell-derived secant moduli were \(\sim 0.1\)--0.3~MPa and explicitly noted to be about an order of magnitude lower at comparable strains, reflecting differences in geometry preservation, boundary conditions, and stress/strain heterogeneity between test modalities \citep{boschetti2012}.

The choice to model collagen fibers in the porcine central cornea as two families of fibers with a weakly anisotropic response, assuming a clear circumferential orientation near the limbus, was made after a careful analysis of the state of the art. At the macroscale, directional tensile testing provides limited evidence for a stable, strongly preferred in-plane axis in the porcine central cornea: in strip extensiometry, the association between anatomical cutting direction and stiffness was reported as statistically insignificant in pigs \citep{elsheikh2008}, and more recent uniaxial/biaxial testing similarly found no significant tensile anisotropy between the superior--inferior (SI) and nasal--temporal (NT) directions \citep{HatamiMarbini2025}. Complementary elastography measurements confirm the presence of anisotropy but suggest it is modest in the central region and that the principal direction may vary between experiments, while the strongest and most consistent directional correspondence emerges near the periphery/limbus \citep{Nguyen2014}. At the microscale, Brillouin microscopy combined with TEM reveals a lamellar \textit{crisscross} architecture with alternating orientations between adjacent lamellae, supporting the interpretation that depth-averaged in-plane alignment in the central stroma can be weak even when local lamellae are individually well organized \citep{Eltony2022}. In contrast, multiple structural studies converge on a more organized peripheral architecture: X-ray scattering mapping classifies the porcine cornea as predominantly circumferential in its preferentially aligned collagen component and highlights the presence of a limbal annulus of aligned collagen \citep{Hayes2007,Meek2009}, a trend that is also consistent with microanatomical imaging showing markedly increased alignment at the corneoscleral limbus compared with the central region \citep{Hammond2020}. Taken together, these observations support a modeling strategy in which the central stroma is represented by a highly dispersed, weakly anisotropic response (implemented here via two symmetric fiber families with dispersion) while allowing a progressive shift toward circumferential organization toward the limbus to reflect the experimentally supported peripheral trend \citep{Hayes2007,Nguyen2014}.

Finally, several factors may influence variability and should be considered when generalizing the results drawn from this study. First, corneal mechanics are strongly geometry dependent; therefore, specimen-specific topography and pachymetry would likely reduce uncertainty and strengthen future inverse identification. Second, testing was performed ex vivo, and despite standardized handling, hydration remains a major confounder in corneal biomechanics. In particular, dextran exposure can dehydrate the cornea and may induce an apparent stiffening effect, which could partially confound treatment-related differences. Accordingly, a sham control group undergoing the same dextran exposure without UV irradiation was not included and should be acknowledged as a limitation. Third, strains were derived from surface kinematics using a membrane-theory formulation, which is practical for DIC-based post-processing but does not explicitly represent through-thickness gradients that may become relevant in some loading regimes. Finally, while the sample size was sufficient to reveal consistent group separation, larger cohorts would enable more precise estimation of inter-specimen variability and more robust assessment of covariate effects.

\section{Conclusion}
Overall, the proposed pipeline provides a practical and scalable basis for analyzing corneal biomechanics using full-field experimental data combined with consistent numerical post-processing. Future studies should test a larger cohort and account for specimen-specific geometry and thickness to reduce uncertainty and improve model fidelity. 
These extensions would further enhance the ability of the framework to quantify how treatment parameters translate into measurable biomechanical changes and to support predictive modeling for refractive procedures and cross-linking optimization.

\section*{Acknowledgments}
This work received funding from the University of Zaragoza (project I-2021/010/PIP OBERON — Opto-Biomechanical Eye Research Network/PIP); from the Spanish Ministerio de Ciencia, Innovación y Universidades through grant PID2023-147987OB-C31, financed by MICIU/AEI/10.13039/501100011033 and by FEDER, EU; and from the Government of Aragon through research group grant T24-20R (co-financed by FEDER). Part of the work was performed at the ICTS ``NANBIOSIS'', specifically the High Performance Computing Unit (U27) of CIBER in Bioengineering, Biomaterials \& Nanomedicine (CIBER-BBN) at the University of Zaragoza. CIBER activities are financed by the Instituto de Salud Carlos III with assistance from the European Regional Development Fund.

\section*{Conflict of interest}
The authors declare that the research was conducted in the absence of any commercial or financial relationships that could be construed as a potential conflict of interest.

\newpage


\end{document}